\documentclass[doublecolumn]{IEEEtran}

\usepackage{stmaryrd}
\usepackage{amsmath}
\usepackage{amssymb}
\usepackage{graphicx}
\usepackage{todonotes}
\usepackage{cite}
\usepackage{algorithmic}
\usepackage{algorithm}
\usepackage{svg}
\usepackage{mathtools}
\usepackage{caption}
\usepackage{subcaption}

\usepackage{float}

\usepackage{comment}
\usepackage{makecell}
\usepackage{xcolor}
\usepackage{amsthm}
\usepackage{epstopdf}
\usepackage{xcolor}
\usepackage{multirow}

\usepackage{bm}
\usepackage{bbm}

\newtheorem{definition}{Definition}

\newcommand{\prob}[1][]{
\ifthenelse{\isempty{#1}}%
      {\ensuremath{P}}%
    {\ensuremath{P\left\(#1\right\)}}%
}
\newcommand{\vect}[1]{\boldsymbol{\mathrm{#1}}}
\newcommand{\mat}[1]{\boldsymbol{\mathrm{#1}}}

\newcommand{\tr}{\mathrm{tr}}

\newcommand*{\inC}[1]{\in\mathbb{C}^{#1}}

\newcommand{\expt}[2]{\mathbb{E}_{#1}\left[#2\right]}
\newcommand{\cn}[2]{\ensuremath{\mathcal{C}\mathcal{N}\left(#1,#2\right)}}
\newcommand{\D}[1]{\mat{D}_{#1}}
\newcommand{\Npath}{N_\textup{path}}
\newcommand{\Nsin}{N_\textup{sin}}

\DeclareMathOperator*{\argmin}{argmin}

\begin{document}

\title{Robust Covariance-Based Activity Detection for Massive Access}

	\author{Jianan~Bai and~Erik~G.~Larsson
	\thanks{
	The authors are with the Department of Electrical Engineering (ISY), Link\"oping University, 58183 Link\"oping, Sweden (email: jianan.bai@liu.se, erik.g.larsson@liu.se). This work was supported in part by ELLIIT,  the KAW foundation, and  the European Union’s Horizon 2020 research and innovation program under grant agreement no.~101013425 (REINDEER).
	The computations were enabled by resources provided by the National Academic Infrastructure for Supercomputing in Sweden (NAISS), partially funded by the Swedish Research Council through grant agreement no. 2022-06725.
}}
	
\maketitle
	
\begin{abstract}
	The wireless channel is undergoing continuous changes, and the block-fading assumption, despite its popularity in theoretical contexts, never holds true in practical scenarios. This discrepancy is particularly critical for user activity detection in grant-free random access, where joint processing across multiple resource blocks is usually undesirable. In this paper, we propose employing a low-dimensional approximation of the channel to capture variations over time and frequency and robustify activity detection algorithms. This approximation entails projecting channel fading vectors onto their principal directions to minimize the approximation order. Through numerical examples, we demonstrate a substantial performance improvement achieved by the resulting activity detection algorithm.
\end{abstract}

\section{Introduction}

In grant-free random access (GFRA) systems, the handshaking process inherent in grant-based schemes is circumvented in order to diminish communication latency.
Nevertheless, due to the open-loop operation of all users, the base station is required to identify active users prior to decoding their payload information. 
The task of activity detection becomes notably challenging when the number of potential users considerably surpasses the channel coherence length, thereby precluding the utilization of mutually orthogonal pilot sequences.

State-of-the-art techniques rely on traffic sporadicity to facilitate accurate activity detection using non-orthogonal pilots. The majority of these methods fall into two primary categories: compressed sensing (CS) and the covariance-based approach. In CS, not only are the identities of active users discerned, but their respective channels can also be recovered. An effective iterative algorithm derived from CS is the approximate message passing (AMP), which has been extensively investigated in studies such as \cite{liu2018massive}. Contrarily, the covariance approach does not directly provide channel estimates, but it can potentially identify a considerably larger number of active users than CS in large, albeit finite, antenna regimes; the scaling law for this approach is formally substantiated in the seminal paper \cite{fengler2021non}.

Predominantly, current schemes restrict to ideal narrowband systems and presume a block-fading channel, an assumption driven by the concept of coherence block and justifiable by the sampling theorem when joint processing or coding across blocks is feasible. Nevertheless, the block-fading assumption may not hold true in the context of activity detection, wherein instantaneous processing is anticipated within a block. It is crucial to acknowledge that the channel continuously experiences significant variations (up to $\pi$ phase shifts) within a coherence block, as the term is nominally defined in the literature, for example in \cite{marzetta2016fundamentals}.

Some efforts have been made to address the channel variations in wideband orthogonal frequency division multiplexing (OFDM) systems. In \cite{jiang2022massive}, subcarriers are partitioned into sub-blocks, and the channel gain is presumed to change linearly within each sub-block; consequently, an AMP-based detection algorithm is developed. The covariance approach is extended in \cite{jiang2022statistical} by assuming independent and identically distributed (i.i.d.) channel taps and capitalizing on the discrete Fourier transform (DFT) structure present in the OFDM symbols.

\textbf{Technical contribution:} We present a unified framework that generalizes the approximation models in \cite{jiang2022massive,jiang2022statistical} from a dimensionality reduction perspective. We propose that the optimal approximation, with respect to minimizing the approximation order, can be achieved by projecting the channel vectors onto their principal directions, a classical outcome derived from principal component analysis (PCA). Our treatment of this subject naturally leads to an extension of the coherence block concept to a statistical perspective, which we term as a prediction horizon. We employ this result to remarkably enhance the robustness of the covariance-based activity detection algorithm.

\textbf{Notations:} Most of our notations follow standard conventions. Particularly, we use $[N]$ to represent the set $\{1,\cdots,N\}$ for any positive integer $N$. We denote $[\vect{x}]_i$ as the $i$-th entry in a vector $\vect{x}$, and $[\mat{X}]_{i,j}$ as the $(i,j)$-th entry in a matrix $\mat{X}$. We use $\D{\vect{x}}$ for a diagonal matrix with $\vect{x}$ on its diagonal.

\section{Signal Model}

We consider a single-cell system comprising a base station with $M$ antennas and $K$ potential users, each with a single antenna. 
Each user $k\in[K]$ is pre-assigned a unique, length-$L$ pilot sequence $\vect{\phi}_k=[\phi_{1k},\cdots,\phi_{Lk}]^T$ that is normalized to have unit energy per symbol, i.e., $\|\vect{\phi}_k\|_2^2=L$. 
We are only interested in the regime where $L\ll K$, and therefore, the pilot sequences of different users are mutually non-orthogonal.
The activity of user $k$ is denoted by a binary variable $a_k$. 

\begin{figure}
	\centering
	\includegraphics[width=4cm]{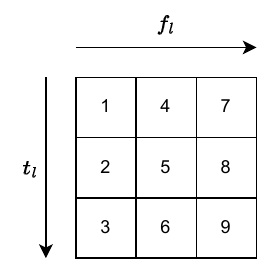}
	\caption{The time-frequency block with $T=F=3$.}
	\label{fig:time-freq-block}
\end{figure}

The active users will transmit their pilot sequences over a time-frequency block of size $T\times F$ (a resource block with $T$ OFDM symbols and $F$ subcarriers). We assume $L=TF$ for simplicity and the time- and frequency indices of the $l$-th pilot dimension are given by
\begin{equation}
\label{eq:index mapping}
	t_l = l - \left\lfloor \frac{l-1}{F} \right\rfloor F \quad\text{and}\quad f_l=\left\lfloor \frac{l-1}{F} \right\rfloor + 1,
\end{equation}
respectively, where $\lfloor\cdot\rfloor$ is the floor function. See Fig. \ref{fig:time-freq-block} for an illustration of such blocks.

Assuming that sufficiently long cyclic prefixes are appended to the OFDM symbols, we obtain $L$ parallel discrete memoryless channels in the frequency domain. Denoting the channel coefficients between user $k$ and antenna $m$ over the pilot dimensions by $\vect{h}_{km} = [h_{1km},\cdots,h_{Lkm}]^T$, the received pilot signals at antenna $m$ is given by
\begin{equation}
	\vect{y}_m = \sum_{k\in[K]} a_k \sqrt{\beta_k} \D{\vect{h}_{km}}\vect{\phi}_k + \vect{w}_m,
\end{equation}
where $\beta_k$ is the received signal strength (the product of the large-scale fading coefficient and the transmit power) of user $k$, and $\vect{w}_m\sim\cn{\vect{0}}{\sigma^2\mat{I}}$ is additive noise.

\section{Dimensionality Reduction of\\Channel Variations}

In the existing literature, the problem of activity detection has been extensively studied under the block-fading model so that the channel coefficients are assumed to be constant during the pilot transmission. This block-fading assumption is motivated by the notion of coherence block/interval (see, for example, \cite{marzetta2016fundamentals}), the size of which depends on the delay spread of multipath propagation and the Doppler frequency resulting from mobility. We note that, however, the wireless channel is always continuously changing, and the block-fading assumption can collapse due to large excess delay and/or high mobility.
The question therefore arises:
\emph{Can we go beyond the block-fading model when designing activity detection algorithms?}

\subsection{A Unified Viewpoint: Dimensionality Reduction}

This question has been partially approached in previous work \cite{jiang2022massive,jiang2022statistical} by either approximating the frequency-selective channel over the subcarriers by a block-wise linear (BWL) model, or leveraging the assumed statistical distribution of the channel impulse response in the delay domain and the DFT structure inherent in the OFDM symbols (hereafter referred to as the DFT-based model). 

We note that both of the models in \cite{jiang2022massive,jiang2022statistical}, as well as the block-fading model, can be interpreted from the perspective of dimensionality reduction and, mathematically, they can be unified into an abstract model
\begin{equation}
\label{eq:channel-approx}
	\vect{h}_{km} \approx \sum_{n\in[N]} \theta_{nkm} \vect{g}_n = \mat{G}\vect{\theta}_{km},
\end{equation}
where $\mat{G}=[\vect{g}_1,\cdots,\vect{g}_N]\inC{L\times N}$ is a deterministic matrix, $\vect{\theta}_{km}=[\theta_{1km},\cdots,\theta_{Nkm}]^T\inC{N}$ is a random vector with i.i.d. entries, and $N$ is the approximation order. 
The interpretation of \eqref{eq:channel-approx} is that the channel vectors $\{\vect{h}_{km}\}$ approximately lie in an $N$-dimensional subspace, $N\ll L$, with the basis $\{\vect{g}_n\}$, and the stochastic nature of $\vect{h}_{km}$ is encapsulated within a significantly reduced representation $\vect{\theta}_{km}$. 

To see how the abstract model \eqref{eq:channel-approx} generalizes those existing models, we first note that, as the simplest example, the block-fading model adopts $N=1$ with $\vect{g}_1=\vect{1}$. The BWL model in \cite{jiang2022massive} divides the subcarriers into $N/2 $ sub-blocks, and each sub-block is associated with two basis vectors, both of which have all-zero entries outside the corresponding sub-block; the first vector's non-zero entries are all ones, accounting for the mean value, while the other features equally spaced entries representing linear variations. The DFT-based model in \cite{jiang2022statistical}, on the other hand, assumes $N$ i.i.d. channel taps in the time domain which results in $N$ basis vectors that are given by corresponding columns in the DFT matrix.

However, to preserve a high approximation accuracy as the channel undergoes more rapid variations, the BWL model necessitates smaller sub-block sizes, which undermines the benefits provided by this model. The DFT-based model fails to exploit the magnitude variation in the power delay profile and the statistical correlation across channel taps due to pulse shaping. As the number of sampled channel taps increases, the algorithm developed based on the DFT-based model may experience escalating complexity and degraded performance. Moreover, these schemes consider only the channel variations across frequency, disregarding the variations over time. This limitation renders these schemes unsuitable for cases requiring multiple (OFDM) symbols due to limited bandwidth.

\subsection{Our Approach: A Principal Component Analysis}

To prepare our answer to the question about extending the block-fading model, we first generalize the notion of coherence interval, which is nominally defined in a deterministic sense, to a statistical viewpoint. The motivation is that even if the channel varies substantially within a block, the fading coefficients may still exhibit strong statistical correlations that can be effectively exploited. Notably, this type of blocks, which we term as a \emph{prediction horizon}, can have a size that is significantly larger than the nominal coherence interval. A more formal definition is provided as follows:
\begin{definition}
	A time-frequency block of size $L$ is an $\epsilon$-approximate prediction horizon of order $N$ if there exist a deterministic matrix $\mat{G}\inC{L\times N}$ and a random vector $\vect{\theta}\inC{N}$ such that the small-scale fading vector $\vect{h}\inC{L}$ can be approximated by $\mat{G}\vect{\theta}$ with approximation error
	\begin{equation}
		\expt{}{\|\vect{h}-\mat{G}\vect{\theta} \|_2} \leq \epsilon.
	\end{equation}
\end{definition}

For any given block size $L$ and approximation order $N$, the optimal choice of $\mat{G}$ that minimizes the approximation error can be obtained by performing a PCA. Specifically, denoting the covariance matrix of $\vect{h}_{km}$ as $\mat{R}$ and its best rank-$N$ approximation (in terms of both spectral- and Frobenius norm) as $\mat{U}\D{\vect{\rho}}\mat{U}^H$, where $\vect{\rho}=[\rho_1,\cdots,\rho_N]^T$ contains the $N$ largest eigenvalues and $\mat{U}=[\vect{u}_1,\cdots,\vect{u}_N]$ comprises of the corresponding eigenvectors, the optimal choice is given by
\begin{equation}
	\mat{G} = \mat{U}\D{\vect{\rho}}^{\frac{1}{2}}.
\end{equation}
By restricting our discussion to Rayleigh fading channels, the corresponding random vector is $\vect{\theta}_{km}\sim\cn{\vect{0}}{\vect{I}_N}$.

To establish a connection with other models from the perspective of PCA, we observe that the block-fading model assumes $\mat{R}=\vect{1}\vect{1}^T$, the BWL model presumes a block-diagonal $\mat{R}$ with a specific structure in each diagonal block, and the DFT-based model assumes $\mat{R}=\mat{F}\mat{F}^H$, where $\mat{F}$ represents the corresponding DFT matrix. In contrast, our approach entails a more efficient exploitation of the covariance information, leading to a more potent dimensionality reduction.

Lastly, we note that the selection of $\mat{G}$ relies on the channel covariance $\mat{R}$, which may not be perfectly known and must be learned from the environment. Fortunately, although the degrees of freedom in a Hermitian matrix generally amount to $L(L+1)/2$, the underlying physical or statistical model typically results in significantly fewer degrees of freedom in $\mat{R}$. For instance, in a wide-sense stationary and uncorrelated scattering (WSSUS) channel, $\mat{R}$ exhibits a Toeplitz-block-Toeplitz structure. The number of unknowns can be further reduced if the underlying physical model is known (with the DFT-based model serving as an extreme case).

\section{What Approximation Order?}

\begin{figure}
	\centering 
	\begin{subfigure}[b]{0.24\textwidth}
		\centering
		\includegraphics[width=\textwidth]{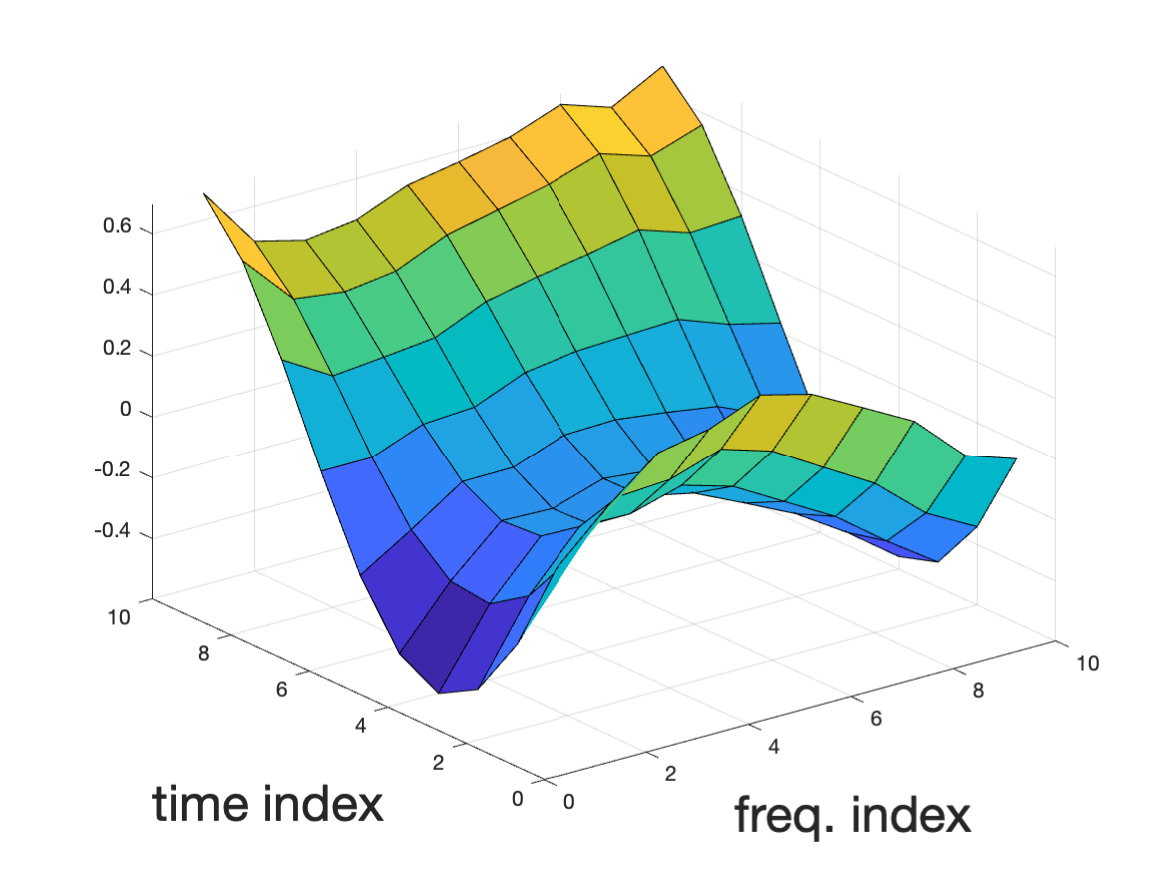}
		\caption{real}
		\label{fig: result L=40}
	\end{subfigure}
	\hfill
	\begin{subfigure}[b]{0.24\textwidth}
		\centering
		\includegraphics[width=\textwidth]{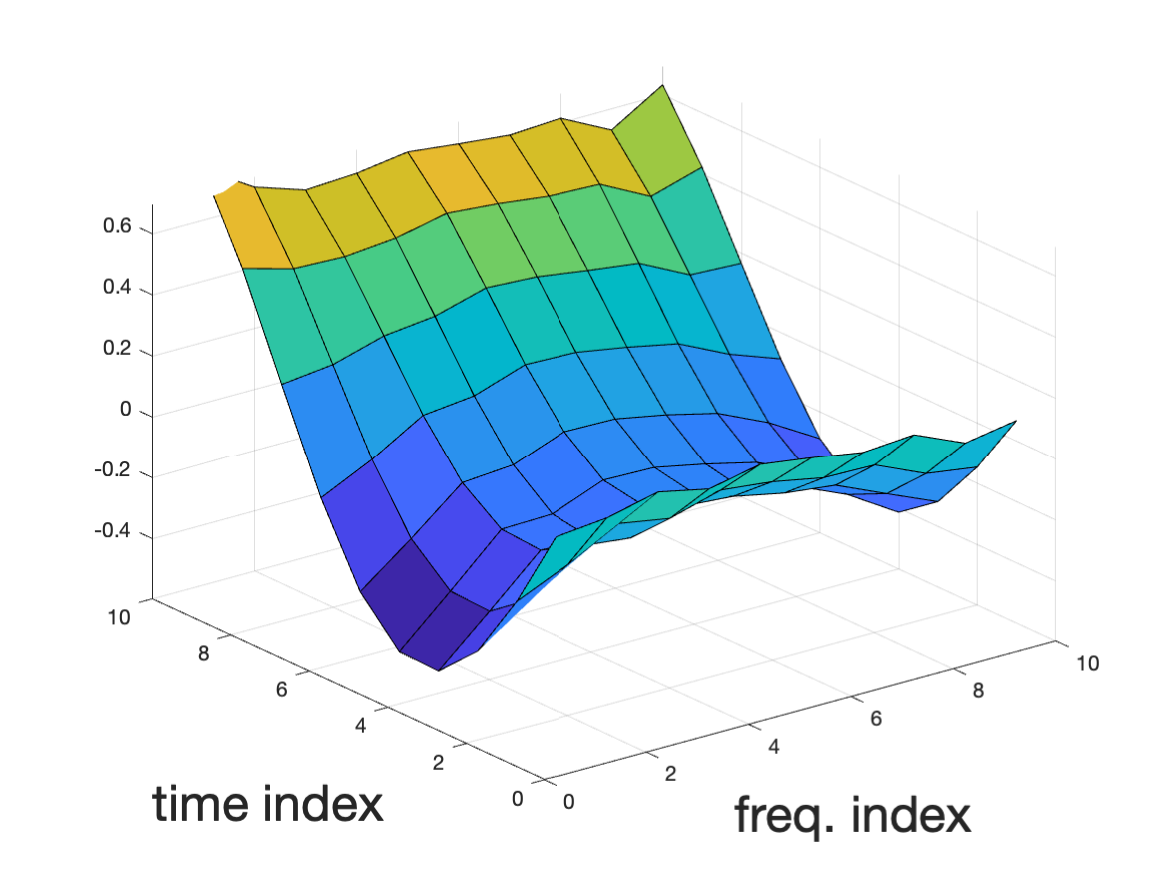}
		\caption{real, approx.}
		\label{fig: result L=20}
	\end{subfigure}
	\hfill
	\begin{subfigure}[b]{0.24\textwidth}
		\centering
		\includegraphics[width=\textwidth]{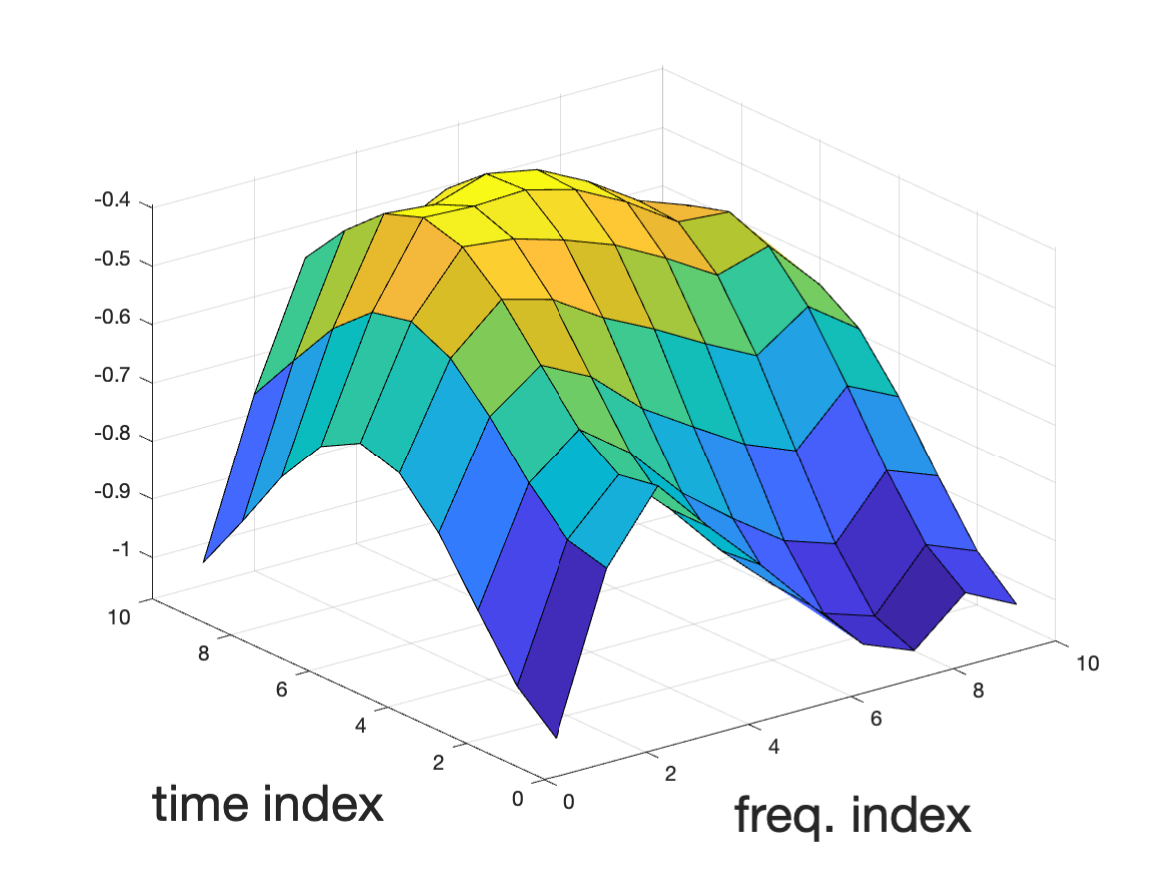}
		\caption{imag}
		\label{fig: result L=40}
	\end{subfigure}
	\hfill
	\begin{subfigure}[b]{0.24\textwidth}
		\centering
		\includegraphics[width=\textwidth]{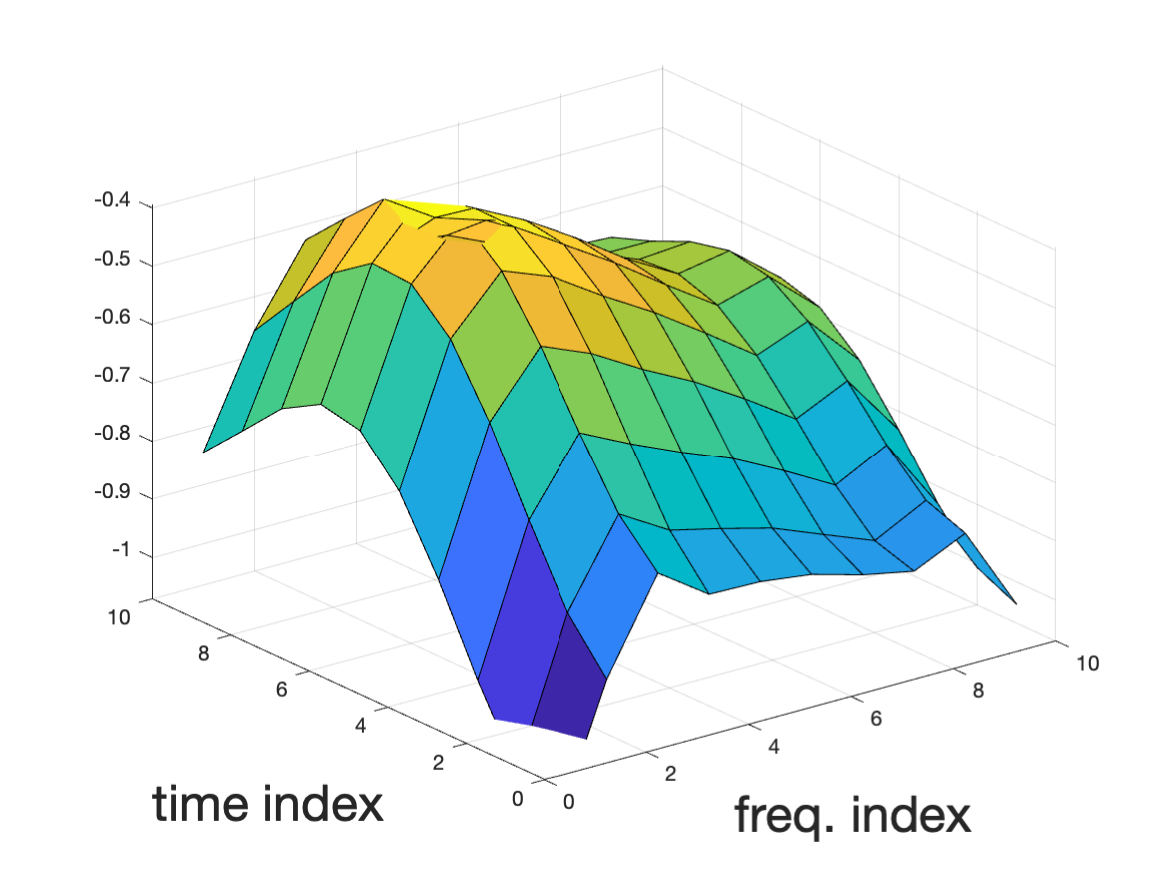}
		\caption{imag, approx.}
		\label{fig: result L=20}
	\end{subfigure}
	\caption{Channel visualization and its 4-order approximation.}
	\label{fig: channel-blcok}
\end{figure}

\begin{figure}
	\centering
	\begin{subfigure}[b]{0.24\textwidth}
		\centering
		\includegraphics[width=\textwidth]{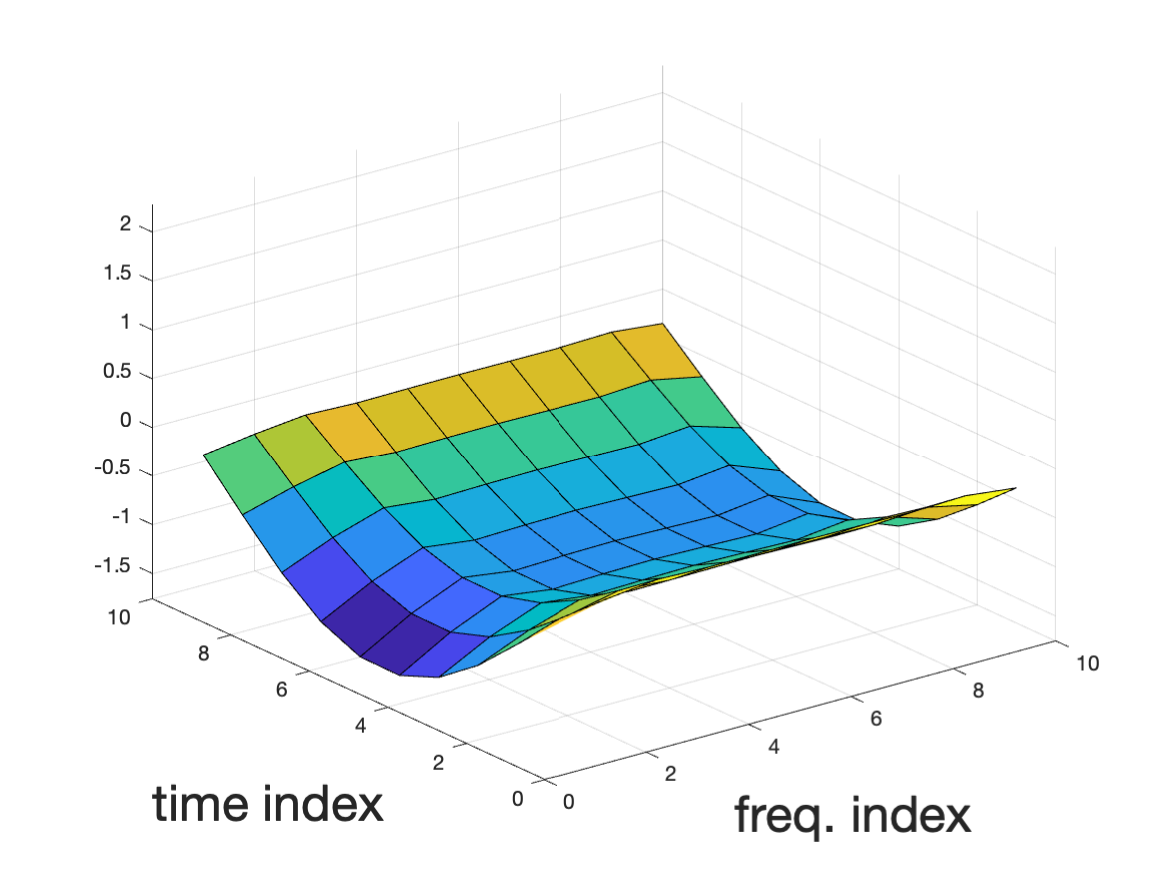}
		\caption{$\vect{g}_1$, real}
		\label{fig: result L=40}
	\end{subfigure}
	\hfill
	\begin{subfigure}[b]{0.24\textwidth}
		\centering
		\includegraphics[width=\textwidth]{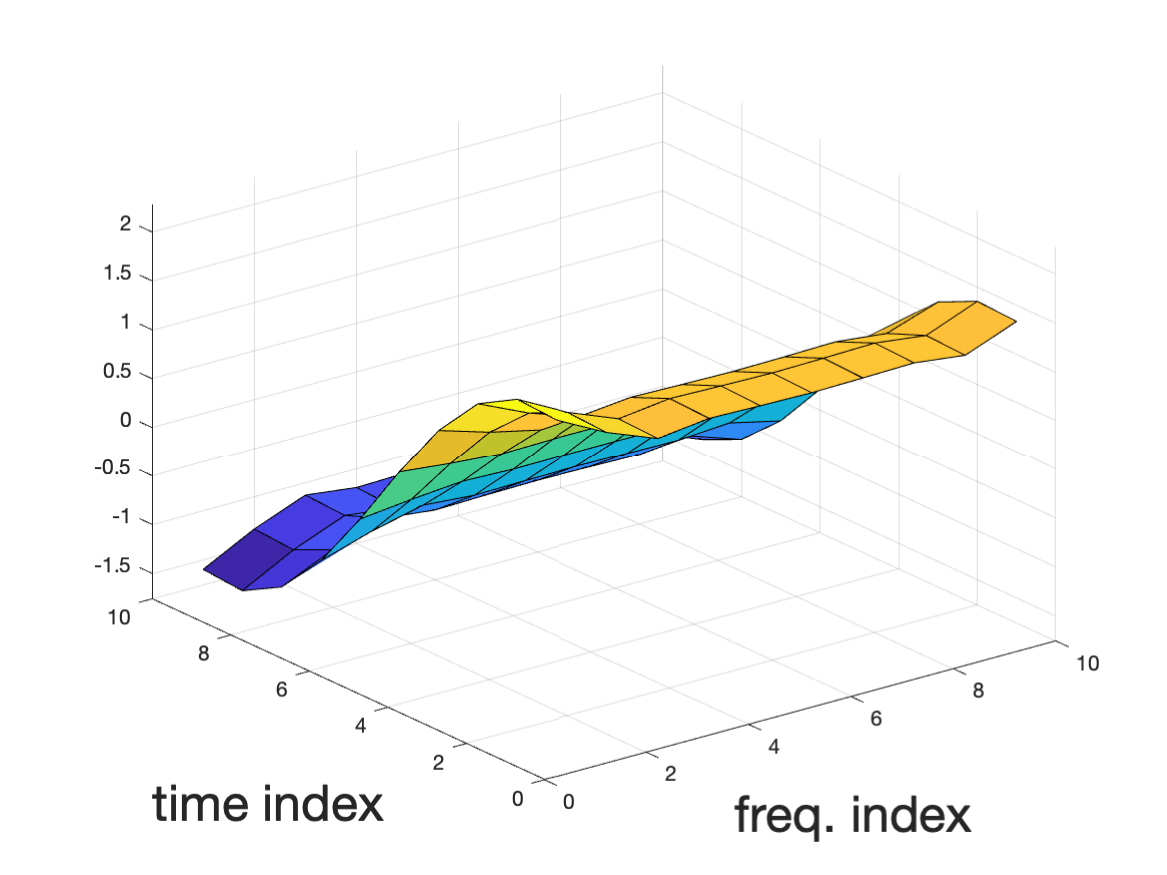}
		\caption{$\vect{g}_2$, real}
		\label{fig: result L=20}
	\end{subfigure}
	\hfill
	\begin{subfigure}[b]{0.24\textwidth}
		\centering
		\includegraphics[width=\textwidth]{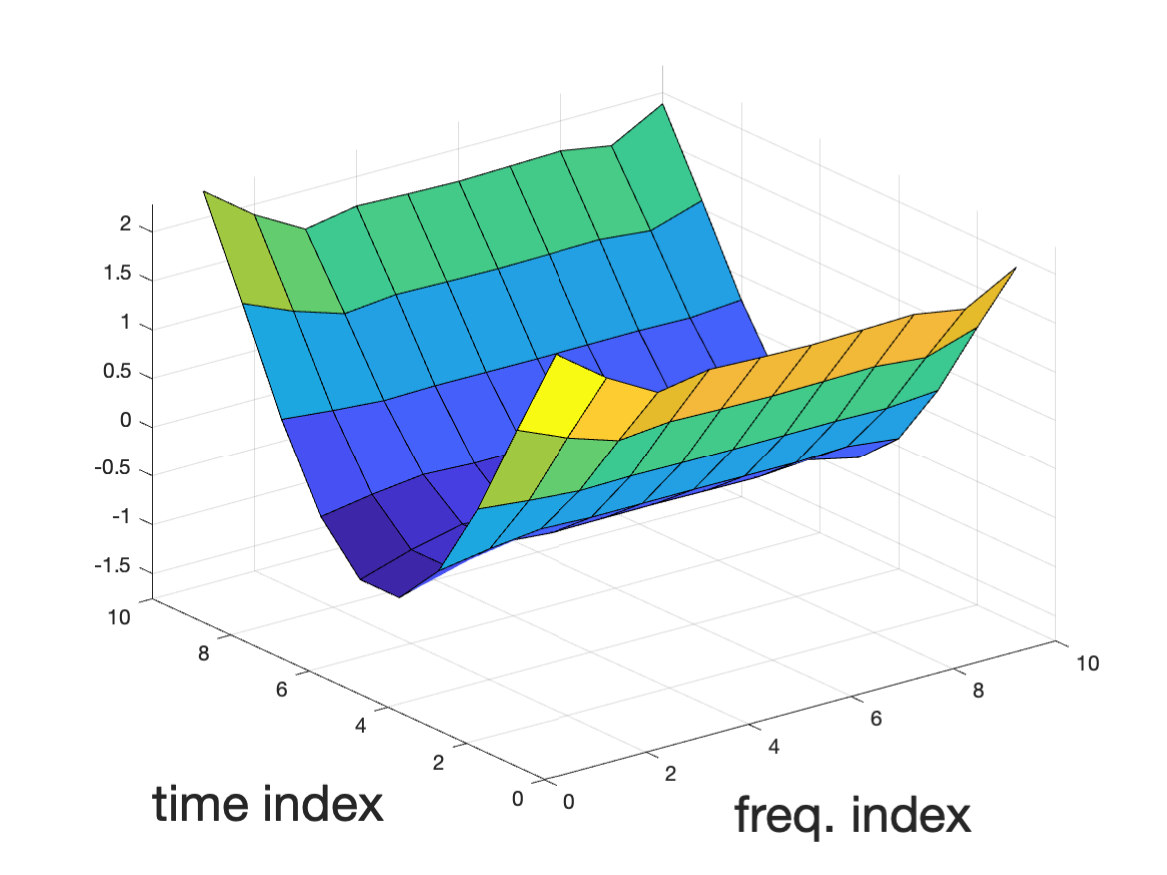}
		\caption{$\vect{g}_3$, real}
		\label{fig: result L=20}
	\end{subfigure}
	\hfill
	\begin{subfigure}[b]{0.24\textwidth}
		\centering
		\includegraphics[width=\textwidth]{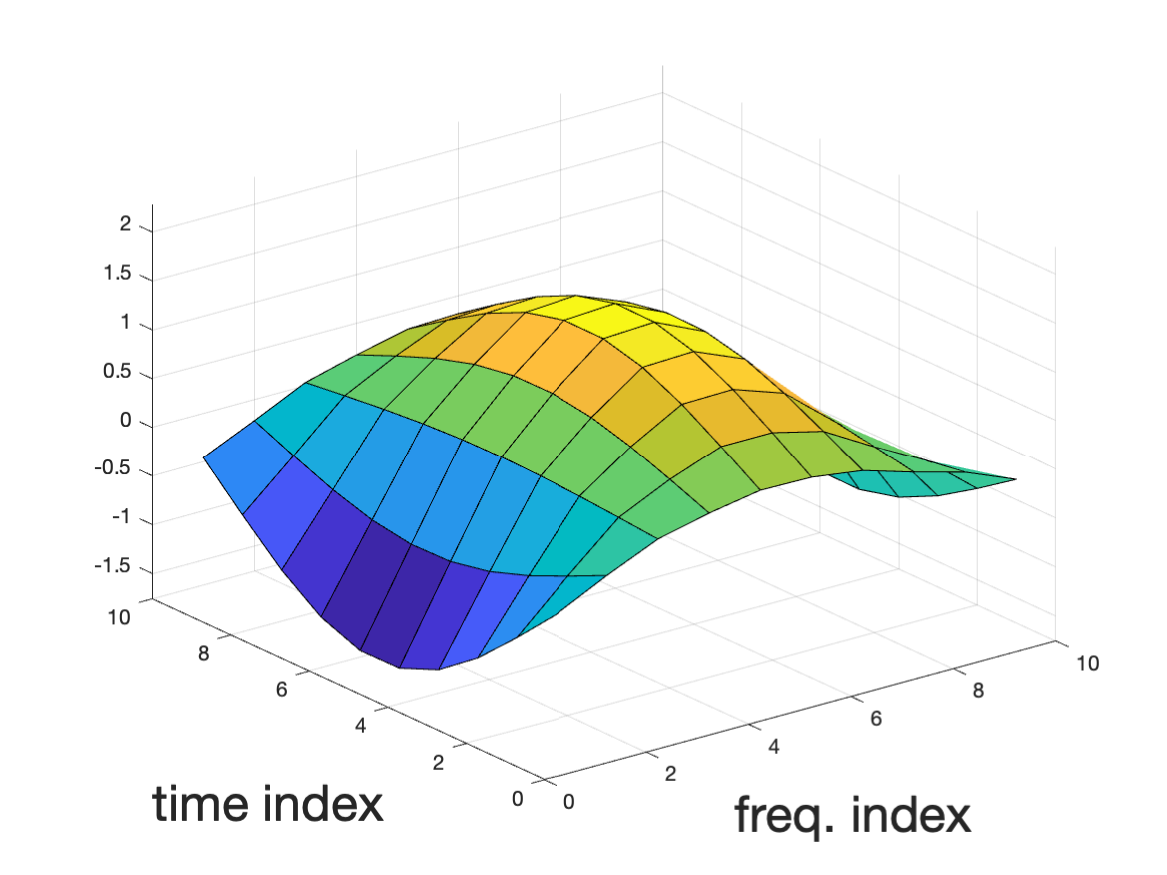}
		\caption{$\vect{g}_4$, real}
		\label{fig: result L=20}
	\end{subfigure}
	\caption{Visualization of the basis vectors.}
	\label{fig:basis-block}
\end{figure}

Our PCA-based model is designed to accurately capture channel variations within a resource block, utilizing the lowest possible approximation order to enable efficient signal processing for activity detection. We observe that even for a physical channel characterized by a large excess delay and high mobility, a low approximation order is sufficient. In the following section, we substantiate this claim through numerical examples.

\subsection{Generation of Channel}

We first generate the impulse response of a WSSUS channel by using the improved sum-of-sinusoids method proposed in \cite{xiao2006novel}. Since the same procedure will be repeated independently for each user-antenna pair, we ignore the subscript ``$km$'' throughout this subsection. For a channel with $\Npath$ paths at different delays, we generate each path independently, and the impulse response of the $i$-th path is given by
\begin{equation}
	q_i(t) = \frac{1}{\sqrt{\Nsin}} \sum_{n\in[\Nsin]} e^{j(\omega_d t \cos\alpha_n + \psi_n)}
\end{equation}
with 
$
	\alpha_n = \frac{2\pi n + \zeta_n}{\Nsin}, n\in[\Nsin],
$
where $\Nsin$ is the number of sinusoids, $\omega_d$ is the maximum Doppler frequency in radians, $\psi_n$ and $\zeta_n$ are i.i.d. distributed over $[-\pi,\pi)$.

We denote the sampling rate by $B$ which is identical to the system bandwidth, and the impulse response of the pulse shaping filter as $p(\cdot)$.
The discrete-time impulse response of the multipath-fading channel is given by
\begin{equation}
	q_{t_l \ell} = \sum_{i\in[\Npath]} \sqrt{c_i} q_i\left(\frac{t_l-1}{B}\right) p(\ell-B\tau_i),
\end{equation}
where $t_l\in[T]$, the fractional power $c_i$ and the delay $\tau_i$ of different paths are determined by a power delay profile defined in, for example, \cite{3gppTR1}. Here, $\ell$ is the time-lag index, which is an integer ranging from $\ell_{\min}$ to $\ell_{\max}$. The smallest time-lag $\ell_{\min}$ is a negative integer, e.g., $-3$, and $\ell_{\max}=\lceil B\tau_{\textup{exc}} \rceil - \ell_{\min}$ with $\tau_{\textup{exc}}=\max_{i\in[\Npath]}\tau_i$ being the maximum excess delay. For each $t_l$, we apply the discrete Fourier transform (DFT) to $\vect{x}_{t_l}=[q_{t_l0},\cdots,q_{t_l\ell_{\max}},q_{t_l\ell_{\min}},\cdots,q_{t_l,-1}]^T$ to obtain the frequency response at $f_l\in[F]$ as
\begin{equation}
	Q_{t_lf_l} = \sum_{i\in[\ell_{\max}-\ell_{\min}+1]} [\vect{x}_{t_l}]_i e^{-j\frac{2\pi (i-1) (f_l-1)}{\ell_{\max}-\ell_{\min}+1}}.
\end{equation}
These coefficients are re-arranged into the length-$L$ channel vector where the $l$-th element, $h_l$, equals to $Q_{t_lf_l}$ with the invertible mapping between $l$ and $(t_l,f_l)$ defined in \eqref{eq:index mapping}.

\subsection{Numerical Example}
\label{sec:channel}

\begin{figure}
	\centering
	\includegraphics[width=6cm]{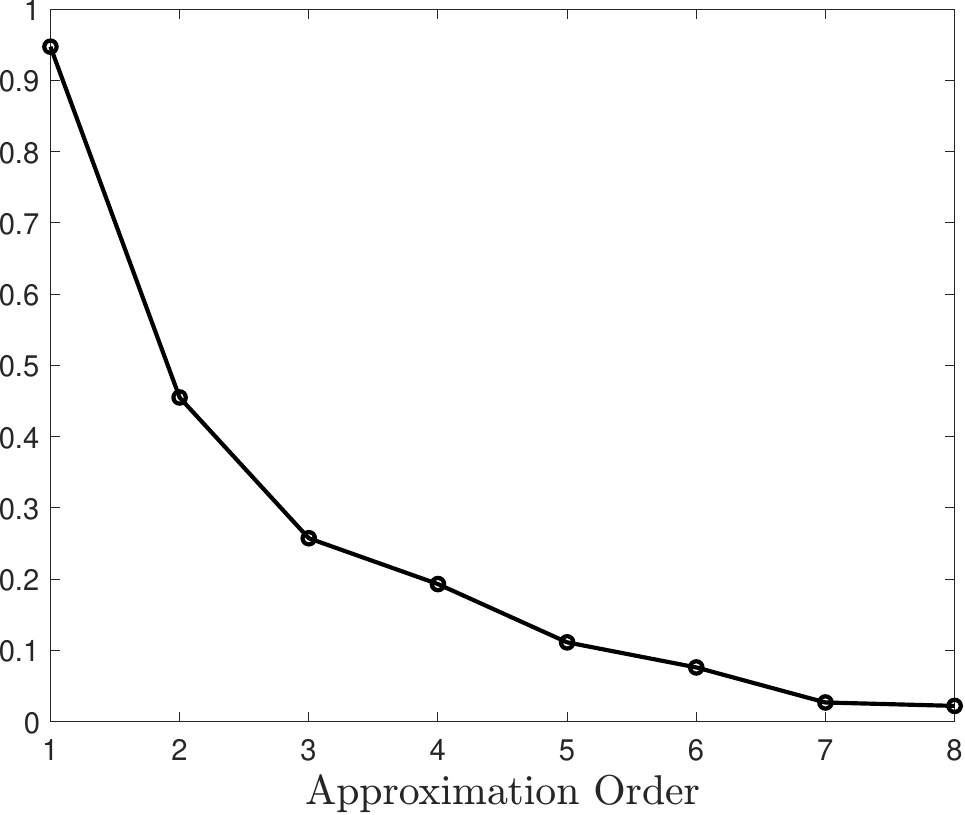}
	\caption{The value of $\kappa$ with different approximation order $N$.}
	\label{fig:kappa}
\end{figure}

We proceed to evaluate the accuracy of the proposed low-dimensional approximation. The time-frequency block size is set to $T=F=10$, resulting in a channel dimension of $L=100$. We employ the Hilly Terrain channel model defined in \cite{3gppTR1}, with a maximum excess delay of $\tau_{\textup{exc}}\approx 18.02$ microseconds and a mobile speed of $120$ km/h. The carrier frequency is set to $3.5$ GHz, which corresponds to a maximum Doppler frequency of $\omega_d \approx 2445.2$ rad/s. The subcarrier spacing is set to $5$ kHz (aligning with the long preamble sequence in the 5G New Radio PRACH formats \cite{lin20195g}), resulting in a system bandwidth for pilot transmission of $50$ kHz. The number of sinusoids in the channel simulator is $\Nsin=20$. For pulse shaping, we utilize a root-raised-cosine (RRC) filter with a rolloff factor of $0.22$ and a symbol rate of $1$.

We perform the eigenvalue decomposition on the sample covariance of the generated fading coefficients to obtain the matrix $\mat{G}$ whose columns are the basis vectors $\{\vect{g}_n\}$.
For an arbitrary user-antenna pair, we obtain the least-squares estimate of the effective channel $\widehat{\vect{\theta}}_{km}$ by minimizing $\|\vect{h}_{km} - \mat{G}\widehat{\vect{\theta}}_{km} \|_2$.
The estimated channel is given by $\widehat{\vect{h}}_{km} = \mat{G}\widehat{\vect{\theta}}_{km}$.
An instance of the channel realization and its order-4 approximation on the time-frequency block are depicted in Fig. \ref{fig: channel-blcok}. The basis vectors ${\vect{g}_n}$ are visualized in Fig. \ref{fig:basis-block} with proper normalization.

To quantify the approximation accuracy, we define $\mat{H} = [\cdots,\vect{h}_{km},\cdots]$ as an $L\times KM$ matrix consisting of the fading coefficients of all user-antenna pairs, $\widehat{\mat{H}} = [\cdots,\widehat{\vect{h}}_{km},\cdots]$ as the channels obtained by the low-dimensional approximation, and $\overline{\mat{H}} = [\cdots,\overline{\vect{h}}_{km},\cdots]$ with $\overline{\vect{h}}_{km} = (\frac{1}{L}\sum_{l\in[L]}h_{lkm})\vect{1}$ as the block-fading approximation. We define the metric
$
	\kappa = \frac{\|\widehat{\mat{H}} - \mat{H} \|_F}{\|\overline{\mat{H}} - \mat{H} \|_F}
$
as the relative approximation error compared with the block-fading model. The value of $\kappa$ for different approximation order $N$ is depicted in Fig. \ref{fig:kappa}. One can observe that a relatively low approximation order, $3\leq N \leq 5$, gives a significantly improved approximation accuracy.

\section{Activity Detection}

By using the abstract channel model in \eqref{eq:channel-approx}, the received pilot signals at antenna $m$ can be re-written as
\begin{equation}
\begin{aligned}
	\vect{y}_m 
	=& \sum_{k\in[K]} a_k \sqrt{\beta_k}\mat{S}_k\vect{\theta}_{km} + \vect{w}_m,
\end{aligned}
\end{equation}
where $\mat{S}_k = [\vect{s}_{1k},\cdots,\vect{s}_{Nk}]$ with $\vect{s}_{nk}=\D{\vect{g}_n}\vect{\phi}_k$ being the effective pilot. 
We note that $\vect{y}_m$ has distribution $\cn{\vect{0}}{\mat{\Sigma}}$ with the covariance matrix
\begin{equation}
\label{eq:cov}
	\mat{\Sigma} = \expt{}{\vect{y}_m\vect{y}_m^H} = \sum_{k\in[K]} a_k \beta_k \mat{S}_k\mat{S}_k^H + \sigma^2\mat{I}.
\end{equation}

Given the observation in \eqref{eq:cov} that the distribution of the received signals $\{\vect{y}_m\}$ is parameterized by the user activities $\vect{a}=[a_1,\cdots,a_K]^T$, it is reasonable to perform a maximum likelihood (ML) detection. However, the binary constraint $\vect{a}\in\{0,1\}^K$ renders the problem combinatorial such that the complexity grows exponentially with $K$. Following the approach in \cite{fengler2021non}, we define $\vect{\gamma}=[\gamma_1,\cdots,\gamma_K]^T$ with $\gamma_k = a_k\beta_k$ and relax the binary constraint by $\vect{\gamma} \in \Gamma$, where $\Gamma$ represents the non-negative orthant $\{\gamma_k \geq 0\}$ when $\{\beta_k\}$ are unknown, and the box constraint $\{0\leq \gamma_k\leq \beta_k\}$ otherwise.

The ML estimation of $\vect{\gamma}$ can be formulated by 
\begin{equation}
\label{eq:ML problem}
	\vect{\gamma}^* = \argmin_{\vect{\gamma}\in\Gamma} f(\vect{\gamma}),
\end{equation}
where 
\begin{equation}
\label{eq:cost}
	f(\vect{\gamma}) = \log|\mat{\Sigma}_{\vect{\gamma}} | + \tr(\mat{\Sigma}_{\vect{\gamma}}^{-1} \widehat{\mat{\Sigma}})
\end{equation}
is the negative log-likelihood function (up to some rescaling and removal of constant terms) with
\begin{equation}
\label{eq:cov-relaxed}
	\mat{\Sigma}_{\vect{\gamma}}=\sum_{k\in[K]}\gamma_k\mat{S}_k\mat{S}_k^H + \sigma^2\mat{I}
\end{equation}
being the parameterized covariance matrix, and the sample covariance $\widehat{\mat{\Sigma}} = \frac{1}{M}\sum_{m\in[M]}\vect{y}_m\vect{y}_m^H$. We note that the cost in \eqref{eq:cost} is the log-determinant divergence between $\widehat{\mat{\Sigma}}$ and $\mat{\Sigma}_{\vect{\gamma}}$ (up to some constant terms), and the ML formulation can also be interpreted from the covariance matching perspective. 

We employ a coordinate descent algorithm to solve the ML problem \eqref{eq:ML problem}.
Specifically, in each iteration of the algorithm, we pick a coordinate (user) $k$ based on some pre-determined schedule and make the update $\vect{\gamma} \leftarrow \vect{\gamma} + d^*\vect{e}_k$ where $\vect{e}_k$ is the $k$-th standard basis vector of $\mathbb{R}^K$, and
\begin{equation}
\label{eq:cost-coordinate}
	d^* = \argmin_{d\in[-\gamma_k,\infty]} f(\vect{\gamma} + d\vect{e}_k).
\end{equation}
(In case that $\beta_k$ is known, the constraint can be replaced by $d\in[-\gamma_k,\beta_k-\gamma_k]$.) Different from the original work \cite{fengler2021non} where the change of $\gamma_k$ results in a rank-$1$ update in the covariance matrix, we need to deal with a rank-$N$ update according to \eqref{eq:cov-relaxed}. This problem has been successfully solved in \cite{jiang2022statistical}, We briefly present the approach in the following. 

By applying Sylvester's determinant identity and the Woodbury matrix identity, we obtain
\begin{equation}
	\log |\mat{\Sigma}_{\vect{\gamma}}\! +\! d\mat{S}_k\mat{S}_k^H | = \log|\mat{\Sigma}_{\vect{\gamma}}| + \log|\mat{I} + d\mat{\Psi}_k|
\end{equation}
\begin{equation}
	(\mat{\Sigma}_{\vect{\gamma}}\!\! +\! d\mat{S}_k\mat{S}_k^H)^{-1} = \mat{\Sigma}_{\vect{\gamma}}^{-1}\!\! -\! d\mat{\Sigma}_{\vect{\gamma}}^{-1}\mat{S}_k(\mat{I}\! +\! d\mat{\Psi}_k)^{-1}\mat{S}_k^H\mat{\Sigma}_{\vect{\gamma}}^{-1}
\end{equation}
with $\mat{\Psi}_k=\mat{S}_k^H\mat{\Sigma}_{\vect{\gamma}}^{-1}\mat{S}_k$. By writing the eigenvalue decomposition of $\mat{\Psi}_k$ as $\mat{V}_k\D{\vect{\lambda}_k}\!\!\mat{V}_k^H$ with $\vect{\lambda}_k = [\lambda_{k1},\cdots,\lambda_{kN}]^T$, the cost function in \eqref{eq:cost-coordinate} can be rewritten as
\begin{align}
\label{eq:cost-1d}
	f(\vect{\gamma}\!+\!d\vect{e}_k) =& f(\vect{\gamma})\! +\! \log|\mat{I}\! +\! d\mat{D}_{\vect{\lambda}_k}|\! -\! d \tr \left((\mat{I}\! +\! d\mat{D}_{\vect{\lambda}_k})^{-\!1}\mat{\Xi}_k \right)\nonumber\\
	=& f(\vect{\gamma}) \!+\!\! \sum_{n\in[N]}\!\!\left(\log(1\!+\!d\lambda_{kn}) \!-\! \frac{d[\mat{\Xi}_k]_{n,n}}{1\!+\!d\lambda_{kn}} \right)
\end{align}
with $\mat{\Xi}_k = \mat{V}_k^H\mat{S}_k^H\mat{\Sigma}_{\vect{\gamma}}^{-1}\widehat{\mat{\Sigma}}\mat{\Sigma}_{\vect{\gamma}}^{-1}\mat{S}_k\mat{V}_k$, and it has the derivative
\begin{equation}
\label{eq:derivative}
	\frac{\partial}{\partial d} f(\vect{\gamma} + d \vect{e}_k) = \sum_{n=1}^N \left( \frac{\lambda_{kn}}{1 + d\lambda_{kn}} - \frac{[\mat{\Xi}_k]_{n,n}}{(1+d\lambda_{kn})^2} \right).
\end{equation}
The optimal $d^*$ can be obtained by comparing the cost value for all feasible stationary points and boundary points. The stationary points where the derivative in \eqref{eq:derivative} equals zero can be obtained by real-root isolation of a polynomial with real-valued coefficients of order $2N-1$. No explicit formula exists for $N\geq 3$, while efficient algorithms were developed \cite{kobel2016computing}. Alternatively, for large $N$, a one-dimensional search can be employed to minimize the cost \eqref{eq:cost-1d} directly.

\subsection{Numerical Evaluation}

We evaluate the performance of activity detection by simulating a single-cell system where the base station has $M=200$ antennas. There are $K=1000$ potential users with $K_{\textup{act}}=100$ active users that are selected uniformly at random during each block. For simplicity, we assume an ideal channel inversion power control such that the signal-to-noise ratio (SNR) equals to 0 dB for all users, i.e., $\beta_1=\cdots=\beta_K=\sigma^2=1$. We consider a time-frequency block of size $T=10$, $F=14$ with pilot length $L=140$, and the fading coefficients are generated by using the same set of parameters as in Section \ref{sec:channel}. The pilot sequences are generated independently from $\cn{\vect{0}}{\mat{I}_L}$ and then normalized to have unit energy per symbol. We run the coordinate descent algorithm for $10$ iterations, and within each iteration, the coordinates are selected by traversing a randomly permuted list of user indices.

\begin{figure}
	\centering
	\includegraphics[width=7.5cm]{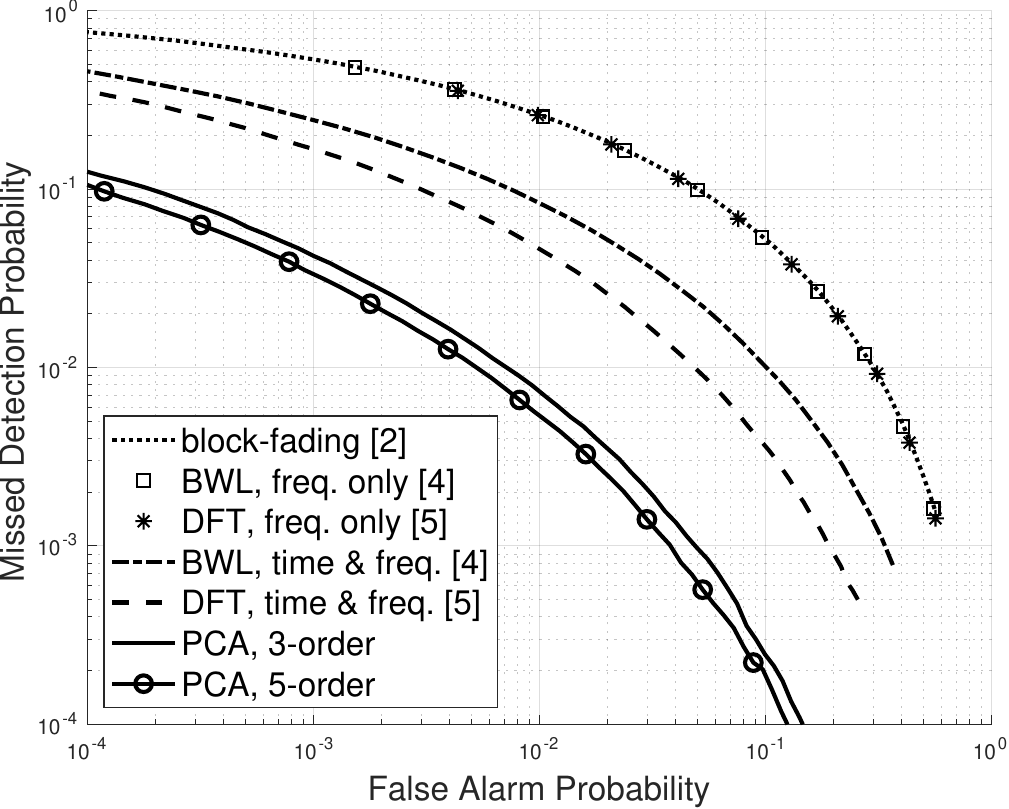}
	\caption{Detection performance.}
	\label{fig:performance}
\end{figure}

In addition to the algorithm for the block-fading model in \cite{fengler2021non}, we also use the BWL model in \cite{jiang2022massive} and the DFT-based model in \cite{jiang2022statistical} as baselines. Since the time variations are not considered in these models, we modify their models for a more fair comparison. In the BWL model, we take the first basis vector as an all-one vector representing the mean value, the second- and third vectors accounting for the linear variations for $t_l\in\{1,\cdots,T/2 \}$ and $t_l\in\{T/2+1,\cdots,T\}$ over time, and the fourth- and fifth vectors for the linear variations for $t_l\in\{1,\cdots,F/2 \}$ and $f_l\in\{F/2+1,\cdots,F\}$ over frequency. In the DFT-based model, the first basis vector is also an all-one vector, while we use the second and last columns in the $T$-point and $F$-point DFT matrices as the remaining basis vectors. We also incorporate the results obtained by ignoring the time variability in these two models (the two basis vectors accounting for time variations are removed). All the basis vectors are properly normalized so that the corresponding random components have unit variance. We note that in \cite{jiang2022massive}, an AMP-based algorithm is designed for the BWL model. Since our main focus is on the channel approximation models instead of comparing different algorithms, we use the adapted covariance approach for this model as well. The detection performance is compared in Fig. \ref{fig:performance}. As one can observe, the PCA-based models significantly robustify the covariance-based activity detection algorithm and outperform the competing models by a considerable margin. The approximation models that ignore the time-variability do not provide an obvious benefit under the highly time-varying environment.

\section{Conclusion}

We generalize the channel approximation models from user activity detection literature into a unified framework following the dimensionality reduction perspective. This naturally leads to an extension of the nominal coherence block concept to a statistical viewpoint which inspires us to exploit the statistical correlation across channel coefficients. Consequently, we propose a PCA-based model to jointly approximate the channel variations over both time and frequency with the lowest possible approximation order. This finding results in an adapted version of the covariance-based activity detection algorithm that is robust under highly varying channel conditions. It is important to note, however, that the accuracy of our PCA-based model depends on the quality of channel covariance information. The acquisition of channel covariance information, while not thoroughly investigated in this paper, requires careful consideration in practice.

\bibliographystyle{ieeetr}
\bibliography{ref}

\end{document}